\documentclass[a4paper]{jpconf} % letterpaper could be used instead of a4paper
\usepackage{graphicx}
\usepackage{float}
\usepackage{mathabx}

\usepackage{amsmath}
\usepackage{orcidlink}
\usepackage{sidecap}
\usepackage{paralist}   
\begin{document}
%You can use \jpcs to insert 'Journal of Physics: Conference Series' in italics
\title{Site Evaluation and Cost Estimation for Cosmic Explorer}

\author{Laurence Datrier\,\orcidlink{0000-0002-0290-3129}, Geoffrey Lovelace, Joshua R. Smith\,\orcidlink{0000-0003-0638-9670}, Andrew Saenz, Amber Romero, for the Cosmic Explorer Project}

\address{The Nicholas and Lee Begovich Center for Gravitational-Wave Physics and Astronomy, California State University, Fullerton, 92831, USA}

\ead{ldatrier@fullerton.edu}

\begin{abstract}
Cosmic Explorer (CE) is a proposed next generation gravitational-wave observatory that would be sited in the United States. As of 2025, CE is in its design phase, with plans to begin operations in the 2030s together with the Einstein Telescope in Europe. CE’s reference design consists of two widely separated L-shaped detectors, one with 20km arms and one with 40km arms, each based on technology proven by the National Science Foundation's highly successful Laser Interferometer Gravitational Wave Observatory (LIGO). There are unique challenges associated with identifying locations suitable for hosting Cosmic Explorer in the conterminous United States, not least of which is the order of magnitude upscaling of the observatory with respect to the 4km LIGO observatories. 
% This paper reviews some aspects of the work towards identifying suitable locations for the Cosmic Explorer observatories. 
Cosmic Explorer’s approach to site evaluation integrates physical, social and cultural criteria. Here we present improvements to the Cosmic Explorer Location Search (CELS) code used to identify and assess locations where CE would have low construction costs incurred by the geology, geography and topography of the land. We also report on efforts to integrate astrophysical requirements established by the Cosmic Explorer Science Traceability Matrix into the site evaluation process. National-level results are presented and combined with results from a related National Suitability Analysis to provide a list of locations that are preliminarily promising for a 40km CE.
\end{abstract}

\section{Introduction}

Cosmic Explorer (CE) is the planned next generation gravitational wave observatory in the United States~\cite{evans2023cosmicexplorersubmissionnsf}. Its reference design consists of two widely separated detectors: one with 40km long arms (CE40), and one with 20km long arms (CE20). The CE Project is, in 2025, in its conceptual/development design and site identification and evaluation phase~\cite{David}. An initial report to the National Science Foundation (NSF) on a preliminary long list of potential site locations for CE is planned for Fall 2026, and a final report on a short list of site locations is anticipated in 2028. The reports will detail candidate locations and synergies between CE and local communities. Criteria for suitable sites for Cosmic Explorer are outlined in \cite{10.1063/5.0242016}. We undertake site evaluation and selection according to the following steps: \begin{inparaenum}[(i)] \item Remote suitability analysis using Geographic Information Systems (GIS) and publicly available data, \item Visits and relationship building~\cite{Joey}, and \item Pending permission, on-site physical and socio-cultural suitability assessments.\end{inparaenum}

The remote evaluation of potential CE sites across the conterminous United States takes in two inputs: the National Suitability Analysis (NSA)~\cite{Warren}, which searches for locations that are conducive to CE's scientific requirements and would offer staff a high quality of life; and the Cosmic Explorer Location Search (CELS), presented here, which evaluates potential configurations based on construction and science factors. 

\section{Cosmic Explorer Location Search (CELS)}

CELS is a python package designed to assist with finding a suitable site for CE~\cite{CELS_codebase}.  It allows the user to estimate costs for the construction of potential detector configurations 
incurred by the topography, geology and geography of the land and is being expanded to take into account the effect of configuration choices on scientific outputs. Much of the code is based on the work outlined in~\cite{Kuns_2020}. 
The ideal CE site is a Euclidean flat plane in 3D space, because the laser beams in CE's arms travel in essentially straight lines. Note that Euclidean flat in 3D space corresponds to a bowl shape in elevation, since the center of the plane is closer to the center of the earth. And since tilting the pendulums that suspend the CE optics couples vertical to horizontal motion, the ideal location has a slope that minimises the average tilt of the suspended mirrors.

In its current form, CELS takes in two data layers: elevation and land cover. CELS currently uses national, public domain datasets as layers. For land cover, we use the 30m resolution National Land Cover Database (NLCD) 2021 dataset from the U.S. Geological Survey (USGS)~\cite{NLCD2021}  The elevation data comes from the elevation layer of the USGS National Map~\cite{NationalMap}. It is a tiled collection of the 3D Elevation Program (3DEP) 1 arc-second resolution data.

\section{Construction costs}

Many factors affect the construction costs, such as the type of land overlapping the observatory's footprint and the elevation along the observatory's arms. We estimate land cover costs by assigning a cost to each type of catalogued land cover, with prohibitively high costs associated with, e.g. water or densely developed areas. 

\begin{figure}[h]
\begin{minipage}{.45\textwidth}
\centering
\includegraphics[width=1\textwidth]{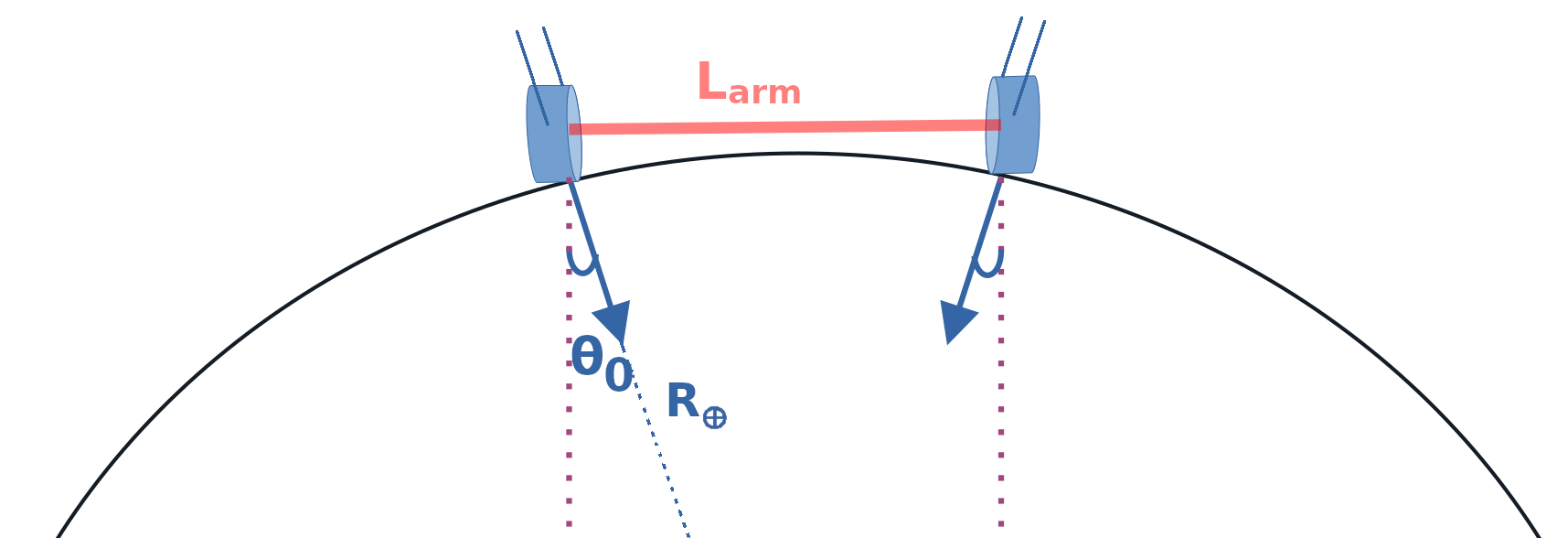}
    \caption{An illustration of the minimum tilt $\theta_{0}$ for a spherical Earth with radius $R$ and an arm length $L_{arm}$. The tilt and curvature are greatly exaggerated.}
    \label{fig:tilt} 
    \end{minipage}
    \hspace{0.1cm}
    \begin{minipage}{.5\textwidth}
    \centering
    \includegraphics[width=1\textwidth]{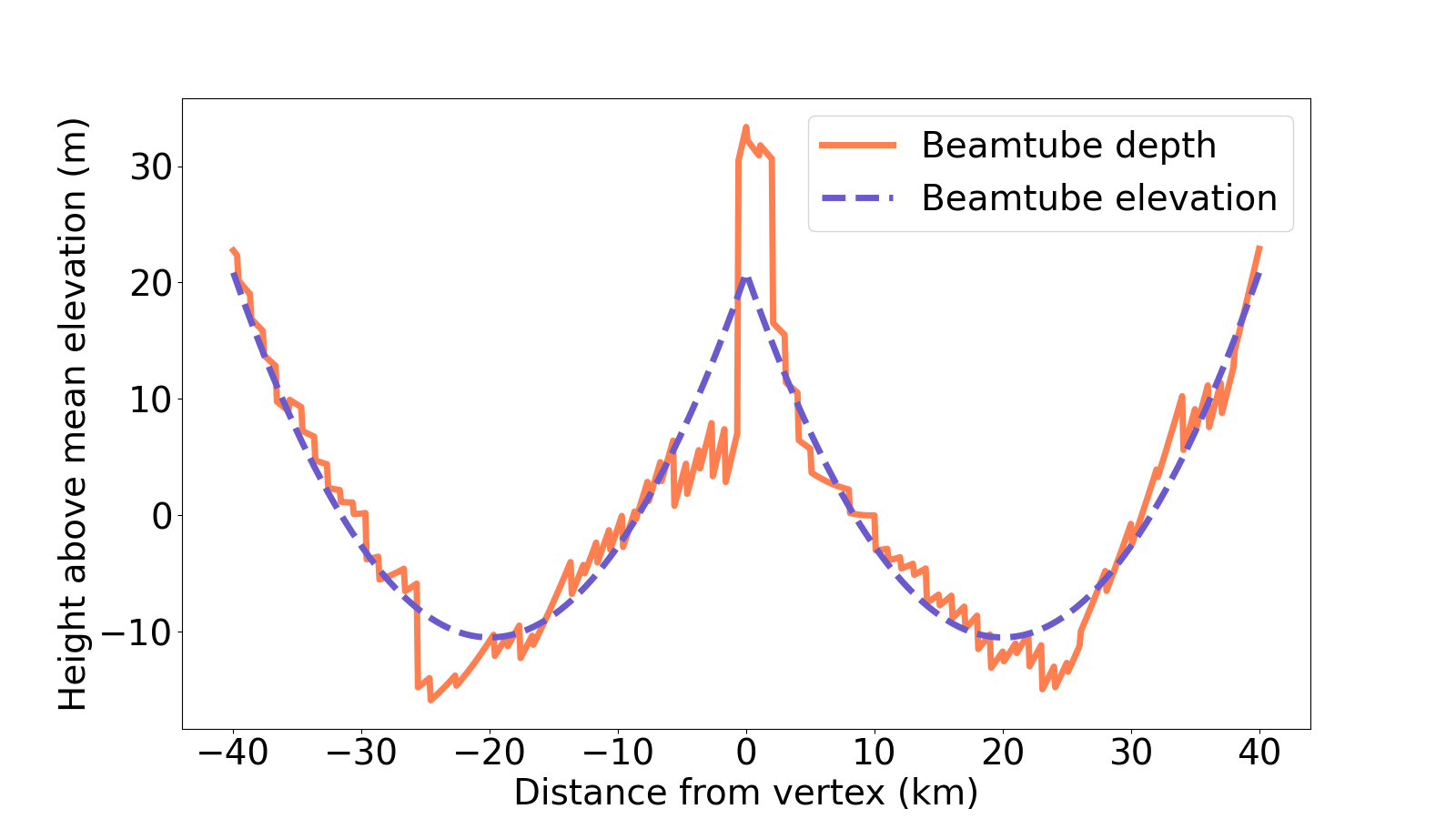}
    \caption{Depth and elevation of the beamtubes, illustrating the bowl-shaped nature of the ideal site.}
    \label{fig:bowl}
    \end{minipage}
\end{figure}

The tilt of a given CE configuration affects its scientific output, rather than its construction cost, but we discuss it here because its evaluation is part of how CELS minimises elevation-related costs. 
Optics suspended from a pendulum follow the local gravitational field at their location, hanging perpendicular to Earth's surface (assuming a spherical Earth). A laser beam between two suspended test masses separated by 40km would not be perpendicular to both. The angle by which the mirrors need to be tilted to be normal to the laser beam for a zero elevation site is defined in CELS as the minimum tilt (Figure~\ref{fig:tilt}), $\theta_{0} = L_{\text{arm}}/2R_{\Earth} \approx 3\text{mrad}$~\cite{Kuns_2020} (with the radius of the Earth $R_{\Earth}$ and an arm length $L_{\text{arm}} = 40$km). CELS then computes tilt scores as 
$C_{\mathrm{tilt}} = 10\left[\left(\theta_{y}/\theta_{0}\right)^{2}+\left(\theta_{y}/\theta_{0}\right)^{2} \right]$~\cite{Kuns_2020},
for a detector tilted in its X- and Y-arms by $\theta_{y}$ and $\theta_{y}$, respectively. 

% In order to build 40km arms (including the vacuum pipes and support structures) that are flat in Euclidean space, trenches and berms need to be built up. 
Constructing flat and level 40km-long arms requires trenches and berms to level any deviations in the land from Euclidean flatness.
We estimate the elevation costs from the volume of material that would need to be excavated or built up (cut and fill), with a cap where digging a tunnel would be cheaper than digging a trench. Specifically, we estimate the cost as $\left(V_{\rm cut} + V_{\rm fill} + |V_{\rm cut} -V_{\rm fill}|\right)/({10~{\rm USD}})$; the first sum corresponds to the amount of earth moved on site, while the absolute value term corresponds to the amount of excess earth that must be brought to or removed from the site. Figure~\ref{fig:bowl} shows, for an arbitrary site, the beamtube depth and elevation above the mean elevation. Work is underway to include soil and rock type. 

\section{Science factors}

Science factors---including tilt, small deviations from a 90$^{\degree}$ arm opening angle, and shorter arm lengths---diminish the scientific returns for a given detector configuration. 

The CE site search is considering opening angles and arm length reductions of 65\textendash115$^{\degree}$ and 2km, respectively---representing a $\sim10$\% strain amplitude penalty. The scientific signal benefit of a detector with an opening angle $\theta$ and an arm length $L$ with respect to one with $\theta=$90$^{\degree}$ and 40km, respectively, can be estimated as: 
$\mathcal{L} = \left(\left(L\sin{\theta}\right)/40\right) ^{2}$. This formula is derived from several science metrics described in~\cite{Read_2023}.
Work is underway to implement this, along with searches that automate varying opening angles and arm lengths. 

\begin{SCfigure}
\centering
\includegraphics[width=25pc]{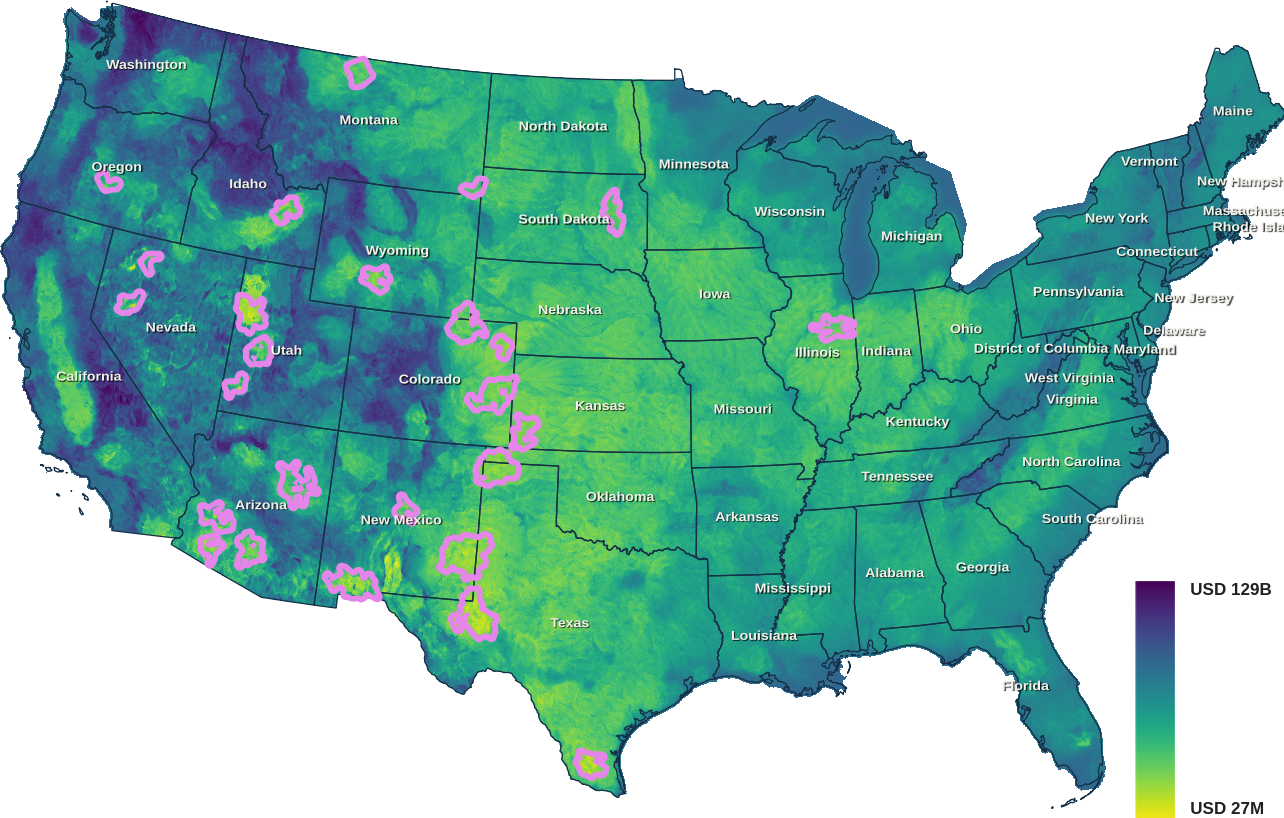}
\hspace{0.1in}
\caption{Results of the national search for 40km CE locations. Each pixel represents the estimated minimum cost over all rotations for a given location (corner station). Yellow represents areas with a lower associated cost and blue represents areas with the highest cost; the color scale is logarithmic. The pink outlines show buffer zones around potential CE configurations for the 26 long-listed sites currently under consideration.\label{fig:CELSmap}}
\end{SCfigure}

\section{Draft locations for Cosmic Explorer}

We have now developed draft long-listed locations for a 40km Cosmic Explorer observatory. Searches for 40km sites are prioritised as per current recommendations for the global gravitational wave detector network to build one CE40 if ET is built, and two CEs (CE40+CE20) if it is not~\cite{KalogeraReport}.
Figure~\ref{fig:CELSmap} shows a map of estimated costs for a 40km CE with a 90$^{\degree}$ opening angle, with outlines indicating 26 locations identified through the NSA~\cite{Warren} and CELS. 

\section{Conclusion and future work}

The broader CE site evaluation team has identified an initial list of 26 suitable locations for a 40km CE. These are still draft locations, and the list may evolve as further searches and visits are carried out. The team expects to complete an initial report to the NSF on these locations by Fall 2026. The final site selection, based on the input from the site evaluation team and other factors, will be made by the NSF. Further CELS work is being carried out on more localised remote evaluation and establishing a 20km long list, in collaboration with the National Suitability Analysis team. Future updates to CELS will incorporate more accurate costing and science factors and deeper integration with the NSA pipeline.

\ack%this is an unnumbered acknowledgement section
The authors are thankful for conversations, feedback and input from the Cosmic Explorer Project and the Cosmic Explorer Consortium--especially from Warren Bristol and Chris Lukinbeal, and Matt Evans and Kevin Kuns.
The authors were supported by the US National Science Foundation grants 2308985 and 2219109 and from Dan Black and Family and Nicholas and Lee Begovich. The Cosmic Explorer site evaluation effort is supported by NSF grants 2308985, 2308986, 2308987, 2308988, 2308989, and 2308990.

%\section*{Appendix}
%Insert Appendix here

\section*{References}
\bibliography{bibli.bib}

\end{document}